\documentclass{aa}
\usepackage{graphicx}
\usepackage{txfonts} 
\begin{document}

\title{Sulfur depletion in dense clouds and circumstellar regions I. H$_2$S ice abundance and UV-photochemical reactions in the H$_2$O-matrix.}
\author{A. Jim\'enez-Escobar \inst{1}, G. M. Mu\~{n}oz Caro \inst{1}} 
\offprints{A. Jim\'enez-Escobar}
\institute{Centro de Astrobiolog\'{\i}a, Dpto. de Astrof\'{\i}sica, INTA-CSIC, Carretera de Ajalvir, 
km 4, Torrej\'on de Ardoz, 28850 Madrid, Spain\\
email: jimenezea@cab.inta-csic.es
}
\date{Received - , 2011; Accepted - , 2011}
\authorrunning{A. Jim\'enez-Escobar et al.}  
\titlerunning{H$_2$S ice abundance and UV-photochemical reactions.} 

  \abstract 
{} 
{This work aims to study the unexplained sulfur depletion observed toward dense clouds and protostars.}
{We made simulation experiments of the UV-photoprocessing and sublimation
of H$_2$S and H$_2$S:H$_2$O ice in dense clouds and circumstellar regions, using the Interstellar Astrochemistry Chamber (ISAC), a state-of-the-art ultra-high-vacuum setup. The ice was monitored in situ by mid-infrared spectroscopy in transmittance. Temperature-programmed desorption (TPD) of the ice was performed using a quadrupole mass spectrometer (QMS) to detect the volatiles desorbing from the ice.}  
{Comparing our laboratory data to infrared observations of protostars we obtained a more accurate upper limit of the abundance of H$_2$S ice toward these objects. We determined the desorption temperature of H$_2$S ice, which depends on the initial H$_2$S:H$_2$O ratio. UV-photoprocessing of H$_2$S:H$_2$O ice led to the formation of several species. Among them, H$_2$S$_2$ was found to photodissociate forming S$_2$ and, by elongation, other species up to S$_8$, which are refractory at room temperature. A large fraction of the missing sulfur in dense clouds and circumstellar regions could thus be polymeric sulfur residing in dust grains.} 
{}

\maketitle

%
%
\section{Introduction}
\label{intro}

Sulfur is depleted in molecular clouds by a factor of 1000 compared to its 
estimated cosmic abundance (Tieftrunk et al. 1994), while in the diffuse 
interstellar medium the abundance of sulfur in the gas phase is comparable to 
the cosmic abundance.
This suggests that there is a form of sulfur in the gas 
phase that was not observed in molecular clouds, which could be atomic sulfur, or alternatively
that sulfur chemistry on icy grain mantles, present in dense clouds and 
regions around YSOs but not in the diffuse interstellar medium, plays an important role. 

There is compelling evidence that supports the role of dust grains on
sulfur chemistry.
Several gas-phase S-containing molecules were observed in hot cores, such as 
OCS, H$_2$S, H$_2$CS, SO, SO$_2$, HCS$^+$, and NS (van der Tak et al. 2003). 
Current gas-phase chemical models are unable to explain the abundances of 
S-species like HCS$^+$ and OCS measured toward protostars (Doty et al. 2004).
The depletion of sulfur is observed not only in dense clouds, but also toward 
Class 0 and Class I sources (Buckle \& Fuller 2003) and toward hot cores (Wakelam et 
al. 2004).
The abundances of S-bearing species, including H$_2$S, suggest that these 
molecules are formed on grain surfaces and subsequently released to the gas 
phase. 

Several S-bearing molecules have been detected in comets. Among them
H$_2$S has the largest abundance, from 0.2 to 1.5\% relative to H$_2$O (Irvine 
et al. 2000). There seems to be a general agreement between the 
molecular abundances observed in circumstellar ices and in comets 
(Bockel\'ee-Morvan et al. 2000). That suggests that H$_2$S is expected to be present in circumstellar ice mantles. The H$_2$S ice will be strongly processed by UV and ion irradiation (Garozzo et al. 2010; Grim \& Greenberg 1987; Moore et al. 2007).

We explore here the possibility that the missing sulfur atoms may be present in icy grain mantles. The cosmic abundance of sulfur is 1.23 $\times$ 10$^{-5}$ N$_{\rm H}$, or 37 times less 
abundant than oxygen (Snow \& Witt 1996), and like oxygen, sulfur belongs to 
group 16 of the periodic table. The electronegativity values of S and O are 2.58 and 
3.44 on the Pauling scale, and the electron affinities are 200 and 141 kJ 
mol$^{-1}$. The bond energy of an S-H bond is 363 kJ mol$^{-1}$, while that of 
an O-H bond is 458.9 kJ mol$^{-1}$, and therefore S-H bond formation is favored
over O-H bond formation.  
Given that H is the most abundant element, S atoms will tend to form H$_2$S molecules because they impinge on icy grain mantles. The feature due to the  \
stretching mode of H$_2$S at 3.925 $\mu$m (2548 cm$^{-1}$) has a band strength 
of $A$ $\approx$ 2.9 $\times$ 10$^{-17}$ cm molecule$^{-1}$ (Smith 1991), and 
$A$(H$_2$O) = 2.0 $\times$ 10$^{-16}$ cm molecule$^{-1}$ (Hagen et al. 1981). 
Therefore, if we make the crude 
assumption that the H$_2$S abundance in the ice is roughly about 1/37 that of 
H$_2$O ice, based on the S/O = 1/37 cosmic abundance ratio, the absorbance 
area of H$_2$S relative to that of H$_2$O might just be 0.4\%,
which is close 
to the detection limit of most observations. With the exception of a weak CH$_3$OH absorption, the feature falls on a relatively {\it clean} part of the mid-infrared 
spectrum. If present, it might be observable in the spectra of 
circumstellar or dense interstellar icy grains. 

Solid H$_2$S has not been detected in the interstellar medium. The presence of 
H$_2$S ice was inferred in W33A, a high-mass protostar (Geballe 1985; 1991), 
and a band at 4.9 $\mu$m (2040 cm$^{-1}$) was attributed to OCS (Geballe 1985, 
Palumbo et al. 1995). The detection of OCS is possible 
since the 4.9 $\mu$m band has a considerable band strength, 
$A$ = 1.5 $\times$10$^{-16}$ cm molecule$^{-1}$ (Hudgins et al. 1993). In addition to OCS, SO$_2$ was detected in ice mantles (Boogert et al. 1997). The H$_2$S detection by Geballe is not fully supported in the literature. Van der
Tak et al. (2003) argue that infrared observations do not support the 
assumption that H$_2$S is the main S reservoir in grain mantles, and they
provide the ISO-SWS observations of W33A as an example.
One of the problems for identifying the 3.925 $\mu$m (2548 cm$^{-1}$) band 
of H$_2$S in H$_2$O-rich ice mantles was the lack of laboratory spectra of 
H$_2$S embedded in an H$_2$O matrix, which is expected to affect this band 
significantly.  

Ultraviolet emission spectra of the coma of comet IRAS-Araki-Alcock showed
the presence of S$_2$, and the spatial profiles indicate a release of this 
species directly from, or very close to, the nucleus (A'Hearn et al. 1983). 
Diatomic sulfur was later found in the comet Hyakutake (Laffont et al. 1996). 
Based on the formation of S$_2$ at 12 K from irradiation of dirty ice
containing H$_2$S, Grim \& Greenberg (1987) suggested 
that S$_2$ was formed in interstellar ice mantles that ultimately aggregate into comets. Subsequently, fast reactions between OCS and metastable S, produced during the dissociation of CS$_2$, were proposed. This supports the formation of S$_2$ in the coma (e.g., A'Hearn et al. 2000). The formation of polymeric sulfur in interstellar environments involves the dissociation of H$_2$S and might, therefore, serve as a probe of energetic processing of the precometary ice.  Ion irradiation experiments of ice analogs containing CO, CH$_3$OH, and S-bearing species lead to formation of OCS and CS$_2$ molecules (Ferrante et al. 2008; Garozzo et al. 2010).

Previously, we performed an experiment consisting of the photoprocessing of 
H$_2$O:CO:NH$_3$:H$_2$S ice followed by warm-up to room temperature. This 
residue was analyzed by means of gas chromatography coupled to mass 
spectroscopy. Among the residue products several N-heterocycles and a number 
of S-bearing molecules were detected. 
S-polymers, S$_6$ through S$_8$, were formed, resulting from the S atoms 
released after photodissociation of H$_2$S ice. But also pentathian 
(S$_5$CH$_2$), hexathiepan (S$_6$CH$_2$), and c-(S-CH$_2$-NH-CH$_2$-NH-CH$_2$) 
were detected (Mu\~noz Caro 2002). The presence of S-containing refractory 
molecules in icy grain mantles, like sulfur polymers, might be the reservoir 
of the missing sulfur in dense clouds and circumstellar environments (Wakelam 
et al. 2005).

To approach the S-depletion dilemma in dense clouds and YSOs, we carried out 
a series of experiments on the deposition of pure H$_2$S or 
H$_2$S in an H$_2$O-matrix under UHV conditions at 
7 K, to mimic interstellar/circumstellar conditions followed by warm-up. 
We measured the mid-infrared spectra of 
both pure H$_2$S ice and H$_2$S in an H$_2$O ice matrix at different 
temperatures from 7 K to sublimation. Similar spectra were reported by Moore 
et al. (2007) at higher temperatures. Using a quadrupole mass 
spectrometer (QMS), we 
obtained the temperature programmed desorption (TPD) plots showing the 
abundances of the molecules released to the gas phase as a function of 
temperature during warm-up. These experiments were repeated, including 
UV irradiation of the ice. The infrared spectra of H$_2$S ice measured in the 
laboratory were compared with spectroscopic observations performed by ISO, 
providing new upper limits on the H$_2$S abundance toward protostars.

The layout of this paper is as follows. In Sect.~2 we describe the 
experimental protocol. The experimental results are presented in Sect.~3. The 
astrophysical implications are discussed in Sect.~4, and the main conclusions summarized in Sect.~5.

\section{Experimental}
\label{expe}
The experiments were performed using the interstellar astrochemistry chamber 
(ISAC). This set-up and the standard experimental protocol were described in 
Mu\~noz Caro et al. (2010). ISAC mainly 
consists of an ultra high vacuum (UHV) chamber, with pressure typically in the 
range P = 2.5--4.0 $\times$ 10$^{-11}$ mbar, where an ice layer made by 
deposition of a gas mixture onto a cold finger at 7 K, achieved by means of a 
closed-cycle helium cryostat, can be UV irradiated. 
 
Samples can be heated in a controlled way from 7 K to room temperature, 
allowing TPD experiments of ices. The 
evolution of the solid sample was monitored by in situ transmittance FTIR 
spectroscopy, while the volatile species were detected by quadrupole mass 
spectroscopy (QMS). The gas line works dynamically, 
and allows the deposition of gas mixtures with the desired composition, which 
is monitored in real time by QMS. A second deposition tube was used for 
codeposition of corrosive gases, such as NH$_3$ or H$_2$S. A prechamber is used to extract the samples while preserving the UHV in the main chamber.

The chemical components used for the experiments described in this paper were 
H$_2$O (liquid), triply distilled, and H$_2$S (gas), Praxair 99.8\%.  H$_2$S 
was deposited through the second deposition tube to prevent it from reacting 
with H$_2$O vapor prior to deposition. For the irradiation experiments, the 
deposited ice
layer was photoprocessed with a microwave-stimulated hydrogen flow discharge 
lamp. The lamp output is $\approx$ 1.5 $\times$ 10$^{15}$ photons s$^{-1}$ 
(Weber \& Greenberg 1985), and the flux was measured at the sample position using
oxygen actinometry is $I_0$ = 2.5 $\times$ 10$^{14}$ photons cm$^{-2}$ s$^{-1}$, 
see Mu\~noz Caro et al. (2010) for details. The emission spectrum 
of the lamp ranges from 7.3 to 10.5 eV (with an average photon energy of 9.2 eV), with main emission at Lyman-$\alpha$ 
(10.2 eV) for a hydrogen pressure $P_{\rm H}$ = 0.5 torr. 

Except for experiment S2 and S6, with resolution 1 cm$ ^{-1}$, the spectral 
resolution was generally 4 cm$ ^{-1}$ to allow acquisition of spectra 
during continuous warm-up. 
The column density of the deposited ice was calculated using the formula\\
\begin{equation}
 N =\int_{band}\frac{\tau_{\nu}d\nu}{A}
\label{column}
\end{equation}
where $N$ is the column density in cm$^{-2}$, $\tau$ the
optical depth of the band, $d\nu$ the wavenumber differential in cm$^{-1}$, and
$A$ the band strength in cm molecule$^{-1}$. The adopted band strength for H$_2$O at 3.05 $\mu$m (3279 cm$^{-1}$) was given in Sect.~\ref{intro}, $A$(H$_2$O) = 2.0 $\times$ 10$^{-16}$ cm molecule$^{-1}$ (Hagen et al. 1981). For H$_2$S, we used $A$(H$_2$S) = 2.9 $\times$ 10$^{-17}$ cm molecule$^{-1}$ (Smith 1991) at 70-80 K. Around 10 K we estimated $A$(H$_2$S) = 2.0 $\times$ 10$^{-17}$ cm molecule$^{-1}$, see Sect.~\ref{noirrad}. The column 
density per number of S atoms in molecule $X$, $N_s(X)$, corresponds to 
$N_s(X)$ = $N(X)$ $\cdot$ $n_S(X)$, where $n_S(X)$ is the number of S atoms in
molecule $X$.

Experiments S7 to S11 involved deposition of the ice layer followed by UV irradiation. The ice mixture compositions were obtained by integrating the infrared absorption bands after deposition. The values in Table~\ref{logIR} correspond to the column density of H$_2$S mixed with H$_2$O ice. They were calculated assuming that the band strength value is equal to that in pure H$_2$S ice. Except for experiment S12, the experiments reported here involved irradiation times shorter than two hours. Calibration experiments of the UV flux using actinometry (Mu\~noz Caro et al. 2010), performed before and after the experiments, showed that the flux remains constant during two hours irradiation. For longer irradiation times of five-hour we found that the flux decreases by 11 \%. Therefore, the UV photon fluence, in photon cm$^{-2}$, is the product of the UV flux value given above, $I_0$, by 
the irradiation time, $t$ in seconds. 

\begin{equation}
 {Fluence} = {I_0 \cdot t}
\label{UV_dose}
\end{equation}

Mason et al. (2006) measure the VUV spectra in the [6.0 - 10.5 eV] range for different molecules in the gas and solid phases. The Ly-$\alpha$ cross section value of water at 25 K obtained from Mason data is 4.0 $\times$ 10$^{-18}$ cm$^2$. The UV absorbed by the ice in our experiments was monitored using
\begin{equation}
UV absorption = \frac{I_0-I}{I_0} = 1 - \frac{I}{I_0} = 1 - \exp{(- \sigma \cdot N(ice))}
\label{Abs_cross}
\end{equation}
where $I_0$ is the above value for the incident UV photon flux at the sample position, $I$ the outgoing UV photon flux, $\sigma$ the UV absorption cross section of the ice, and $N(ice)$ the total ice column density.
The thickest irradiation, experiment S10, was performed with a total ice thickness of 0.25 $\mu$m where 95\% of the impinging photons are absorbed by the ice according to Eq.~\ref{Abs_cross}.For H$_2$S ice, there is no reported $\sigma$ value. We adopt the value of 9.01 $\times$ 10$^{-16}$ cm$^2$nm in the [121.2 - 159.3 nm] range or 6.7 $\times$ 10$^{-18}$ cm$^2$ per photon of 9.2 eV for H$_2$S in the gas phase (Lee et al. 1987). 	
For experiment S8, corresponding to the thickest layer of pure H$_2$S, we find that the UV transmission of H$_2$S ice drops to 51$\%$ for 0.032 $\mu$m.

Experiment S12 differs from the others because it involved simultaneous deposition and irradiation. The gas was condensed at a rate of about 2.5 $\times$ 10$^{14}$ molecules cm$^{-2}$ s$^{-1}$. Because the UV flux is 2.5 $\times$ 10$^{14}$ photons cm$^{-2}$s$^{-1}$, very roughly, the ice was exposed to about one photon per molecule on average. 
For this experiment the column density of the ice was estimated as follows. The H$_2$O:H$_2$S = 13:100 ratio of the gas mixture in the gas line during deposition was measured by QMS and found to remain constant. The deposition rate, in molecules cm$^{-2}$s$^{-1}$, was calibrated from previous experiments, allowing the estimation of the total ice column density for a total deposition time of 252 min. and in particular the value of N(H$_2$S) given in Table~\ref{logIR}.

\section{Experimental results}
\label{results}
Table~\ref{logIR} lists the experimental parameters. The first three columns 
indicate, respectively, the label of the experiment, if the experiment 
involved irradiation, and the starting ice composition. The fourth column gives
the fluence in photons cm$^{-2}$, the fifth column the heating rate in K 
min$^{-1}$. The sixth column is the column density of deposited H$_2$S ice. 
In experiment S12 the column density of H$_2$S was inferred from gas phase composition during deposition and irradiation (QMS data) and calibration with previous experiments. The deposition rate (in molecules cm$^{-2}$ s$^{-1}$) of the ice in that experiment was chosen to be equal to the lamp flux (photons cm$^{-2}$ s$^{-1}$). The seventh column is the column density of HS$_2^\cdot$ after 100 min irradiation, N$_s$(HS$_2^\cdot$) = 2 N(HS$_2^\cdot$), divided by the column density of deposited H$_2$S, N$_s$(H$_2$S) = N(H$_2$S), in percent, obtained from integration of the infrared bands.These are the three experimental points in Fig. 7 (bottom panel) for around 100 min of irradiation. No value of $A$(HS$_2^\cdot$), the band strength of HS$_2^\cdot$, was found in the literature. We assumed $A$(HS$_2^\cdot$) = 0.5 $A$(H$_2$S) because H$_2$S has 2 S-H bonds while HS$_2^\cdot$ only has one. The values obtained are affected by an error of 25 \%. The last column provides an upper 
limit value for the total of products in percent, estimated from the decrease in the 
infrared H$_2$S band at 3.925 $\mu$m (2548 cm$^{-1}$) upon irradiation, i.e. 1-$\frac{N_s(H_2S)}{N_s(H_2S)_i}$ where N$_s$(H$_2$S)$_i$ is the initial column density of H$_2$S in the ice and N$_s$(H$_2$S) the column density after irradiation, assuming that all the photolyzed H$_2$S molecules led to product formation. 

Section.~\ref{noirrad} describes the results from the experiments 
consisting on the deposition and controlled warm-up of pure H$_2$S ice, 
experiments S1 to S4, or H$_2$S in an H$_2$O-matrix,
experiments S5 and S6. Section.~\ref{irrad} describes the 
results from experiments S7 to S9, where the ice was irradiated after 
deposition. Section.~\ref{irrad_2} describes experiments S10 and S11 involving the deposition of H$_2$S:H$_2$O ice mixtures followed by irradiation, and experiment S12, involving simultaneous deposition and irradiation.

\subsection{Annealing of H$_2$S ice experiments}
\label{noirrad}
\begin{figure}  [ht!]
    \centering

    \includegraphics[width=7.5cm]{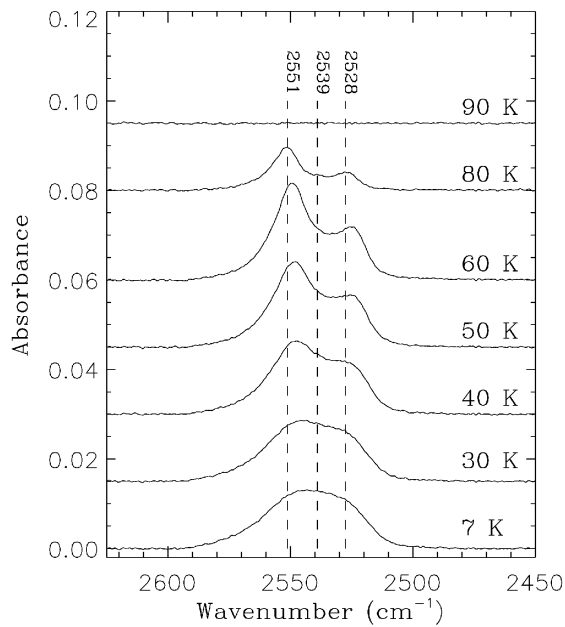}
     \includegraphics[width=9.cm]{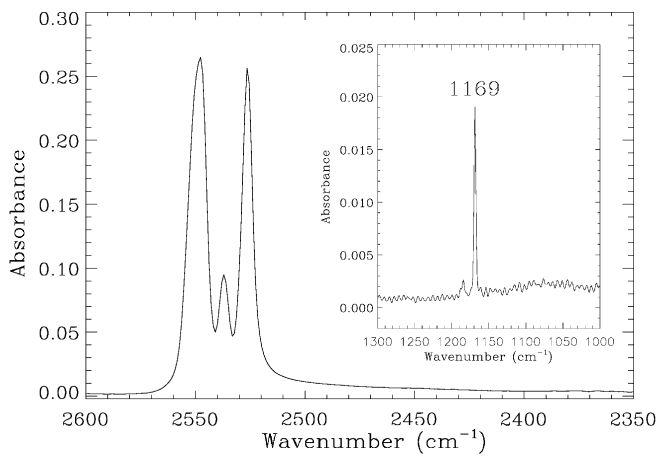}
   
    \caption{Top: Infrared spectra of H$_2$S ice with 1 cm$^{-1}$ spectral 
        resolution at different temperatures 
        during warm-up, corresponding to experiment S2 of Table~\ref{logIR}.
        Bottom: Infrared spectrum showing the stretching modes of crystalline 
        H$_2$S deposited at 70 K. Inlet shows the bending mode absorption.}
       \label{ir_fig}
\end{figure}

The top panel of Fig.~\ref{ir_fig} displays the infrared band of pure H$_2$S at 1 cm$^{-1}$ spectral resolution after deposition 
at 7 K and during warm-up to 90 K, the temperature at which sublimation was 
complete. The band of H$_2$S ice at 7 K, considered to be amorphous, reaches its maximum around 2543 cm$^{-1}$ (3.93 $\mu$m). Crystallization is expected to take place during warm-up and as a result, the different 
vibrational modes become better defined. This is shown as a change in 
the band profile, which is observed around 40 K, leading 
to three subfeatures that were attributed to 2551 cm$^{-1}$ ($\nu_3$, 
antisym. HS-str.), 2528 cm$^{-1}$ ($\nu_1$, sym. HS-str.), and 
2539 cm$^{-1}$, which could be due to the remaining amorphous H$_2$S. 
We performed a similar experiment depositing a thick H$_2$S ice layer to 
detect the weak H-S bending mode. The bottom panel of Fig.~\ref{ir_fig} 
shows the stretching mode of crystalline H$_2$S and the inlet shows the 
bending mode around 1169 cm$^{-1}$ (8.554 $\mu$m). Our spectra are similar 
to those previously published, see Moore et al. (2007) and ref. therein.
 
\begin{figure}[ht!]
    \centering

    \includegraphics[width=9.cm]{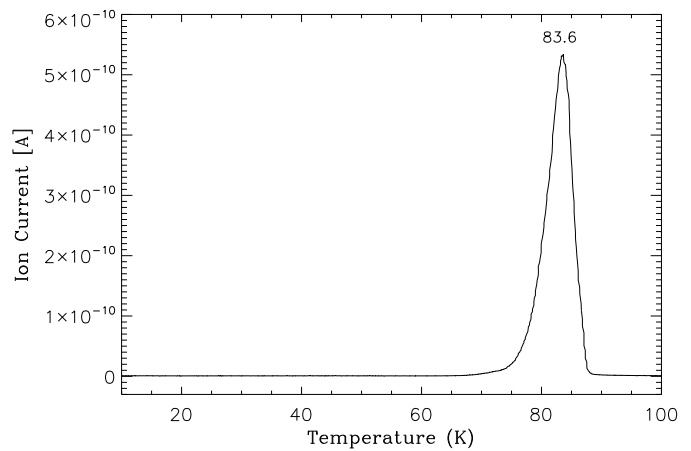}         
        \caption{Thermal desorption of pure H$_2$S, with a heating rate of 
        1.0 K min$^{-1}$, corresponding to experiment S4 of Table~\ref{logIR}. The ion current in Ampere, represented on the y-scale, corresponds roughly to the partial pressure in mbar.}
        \label{qms_fig}
\end{figure}

During sequential warm-up, H$_2$S ice sublimation is also traced by the 
detection of gas phase molecules at the QMS located on the main chamber
of ISAC.
This QMS data, henceforth referred to as the TPD curve, is displayed in 
Fig.~\ref{qms_fig} for m/z = 34. 
Around 70 K a very sharp increase in the TPD curve is observed, see 
Fig.~\ref{qms_fig}. That corresponds to the sublimation stage of the ice.

\begin{figure}[ht!]
    \centering
    \includegraphics[width=8.cm]{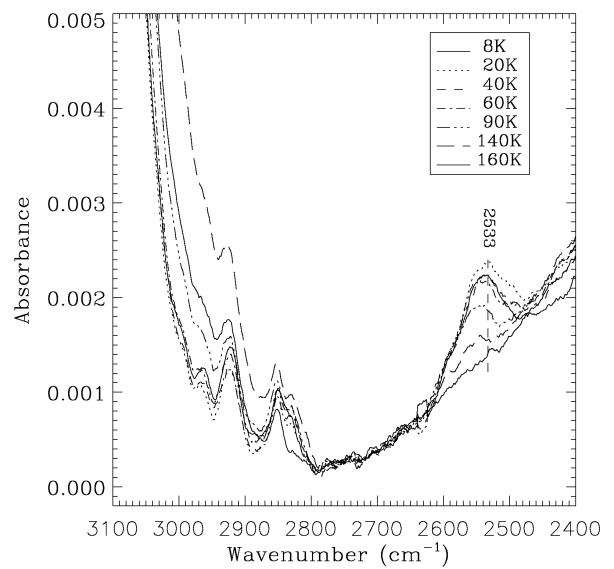} 

        \caption{Infrared spectra of the H$_2$S:H$_2$O = 7.5:100 ice mixture 
        at different temperatures 
        during warm-up, corresponding to experiment S6 of Table~\ref{logIR}.}
        \label{ir_S4}
\end{figure}

\begin{figure}[ht!]
    \centering

    \includegraphics[width=8.5cm]{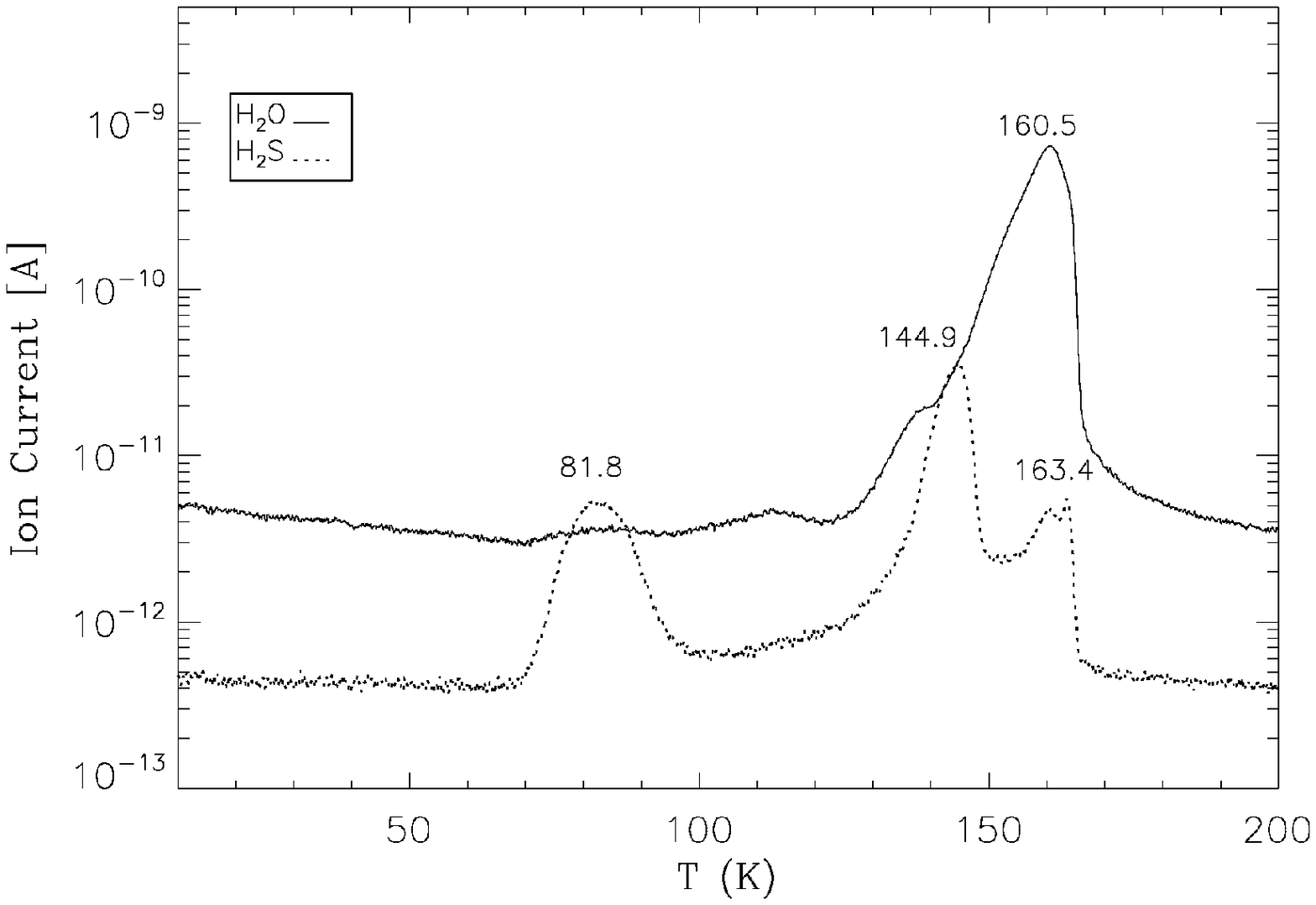} 

        \caption{Thermal desorption of the H$_2$S:H$_2$O = 7.5:100 ice 
         mixture, with a heating rate of 1 K min$^{-1}$, corresponding to 
         experiment S6 of Table~\ref{logIR}. The ion current in Ampere, represented on the y-scale, corresponds roughly to the partial pressure in mbar.}
        \label{qms_S4}
\end{figure}

To study the effect of the H$_2$O-matrix, we deposited an H$_2$S:H$_2$O = 7.5:100 ice mixture, see 
experiment S6 parameters in Table~\ref{logIR}. Figure.~\ref{ir_S4} shows the infrared band of H$_2$S for this experiment at 1 cm$^{-1}$ spectral resolution. The main
effect of the interaction between H$_2$O and H$_2$S molecules in the solid
is the widening of the H$_2$S band (FWHM of 75 cm$^{-1}$  for H$_2$S in an H$_2$O ice matrix compared to 
43 cm$^{-1}$ for pure H$_2$S ice), see Fig.~\ref{ir_S4}. 
The wavenumber position of the H$_2$S infrared band does not change significantly when mixed 
with H$_2$O.
The TPD data collected during this experiment are shown in Fig.~\ref{qms_S4}. 
The TPD curve of H$_2$O for m/z = 18 in Fig.~\ref{qms_S4} is similar to the one of pure H$_2$O ice, which is not shown.
The data show the transitions from amorphous solid water (asw) of low density 
to cubic ice (Ic) at 130--140 K, and from cubic ice (Ic) to hexagonal ice 
(Ih) around 150 K (Hagen et al. 1981).
On the other hand, the H$_2$O matrix has a strong effect on the desorption 
of H$_2$S ice, as inferred from comparison of the TPD curve of pure 
H$_2$S, see Fig.~\ref{qms_fig} for m/z = 34, with the TPD curve of H$_2$S in experiment S6, 
see Fig.~\ref{qms_S4} for m/z = 34. The maximum around 82 K in Fig.~\ref{qms_S4} 
corresponds to the observed peak for pure H$_2$S in Fig.~\ref{qms_fig}. The peaks around 145 and 
163 K were not observed for pure H$_2$S ice, which indicates that a
fraction of the H$_2$S 
molecules are retained at higher temperatures in the
H$_2$O-matrix. The desorption peak of
H$_2$S around 145 K forms during the sublimation of the H$_2$O ice bulk, while 
the less intense peak around 163 K appears right after the maximum in the 
sublimation of H$_2$O around 160 K.

\begin{figure}[ht!]
    \centering
      \begin{tabular}{r}

    \includegraphics[width=9.cm]{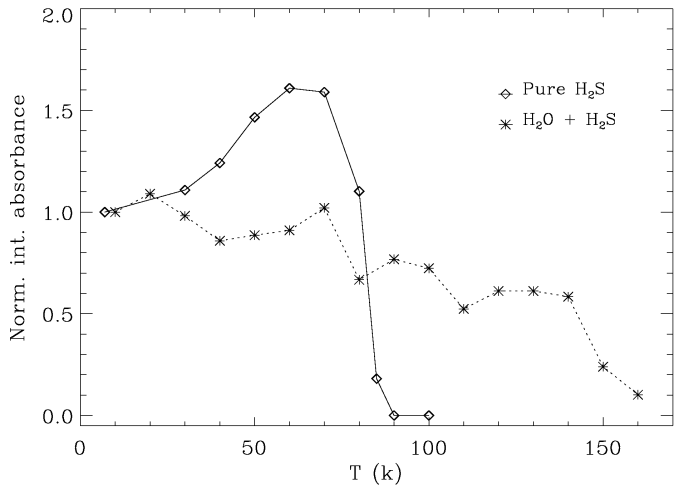} \\     

   \end{tabular}
         
        \caption{Normalized integrated infrared absorbance of pure H$_2$S ice 
        (experiment S3) and H$_2$S:H$_2$O = 7.5:100 ice (experiment S6, see 
        Table~\ref{logIR}) at different temperatures during warm-up.}
       \label{ir_integ}
\end{figure}

The effect of the H$_2$O-matrix on the desorption of H$_2$S ice 
was also monitored by integrating the infrared H$_2$S absorption band for 
pure H$_2$S and for the H$_2$S:H$_2$O = 7.5:100 ice mixture at different 
temperatures during warm-up, see Figs.~\ref{ir_fig} and ~\ref{ir_S4}. The values of the integrated absorbance are shown in Fig.~\ref{ir_integ}. It is important to note that the QMS data presented in Figs.~\ref{qms_fig} and ~\ref{qms_S4} reflects the detection of desorbed molecules in the gas phase, while the integrated infrared absorbances not only show a decrease in the infrared band due to desorption, they are also affected by crystallization. 
First, the main difference between the pure H$_2$S ice and the 
H$_2$S:H$_2$O = 7.5:100 ice experiments is the crystallization process, as seen 
in Fig.~\ref{ir_integ}: while no increase in the integrated absorbance is 
observed in the H$_2$S:H$_2$O = 7.5:100 ice experiment, the high rise of the integrated 
absorbance for pure H$_2$S ice is due to crystallization.
The band strength of $A$(H$_2$S) = 2.9 $\times$ 10$^{-17}$ cm molecule$^{-1}$ commonly used (Smith 1991) refers to H$_2$S at 80 K when full crystallization is attained. We obtained a band strength value at 10 K of $A$(H$_2$S) = 2.0 $\times$ 10$^{-17}$ cm molecule$^{-1}$, using the expression
\begin{equation}
  A_{10K} =\frac{\rm Int.abs_{10K}}{\rm  Int.abs_{70K}}\cdot A_{70K}
\label{band_strength}
\end{equation}
where $A_{10K}$ and $A_{70K}$ are the band strength values of H$_2$S at 10 and 70 K, respectively, Int.abs$_{10K}$ and Int.abs$_{70K}$ are the integrated absorption values of the H$_2$S band at 10 and 70 K.
 
Second, the integrated absorbance of pure H$_2$S ice dropped to zero above 90 K; i.e., all the H$_2$S ice molecules 
have desorbed at that temperature, but in the H$_2$S:H$_2$O = 7.5:100 (experiment S6 of Table 1) ice experiment H$_2$S molecules are kept in the H$_2$O matrix up to 130 K.
The much higher sensitivity of QMS compared to FTIR spectroscopy explains why 
desorption of H$_2$S is observed at temperatures above 130 K by QMS during the same experiment, see
Fig.~\ref{qms_S4} and its interpretation as given above.

\subsection{Irradiation of pure H$_2$S ice experiments}
\label{irrad}
\begin{figure}[ht!]
    \centering
      \begin{tabular}{r}

    \includegraphics[width=7.cm]{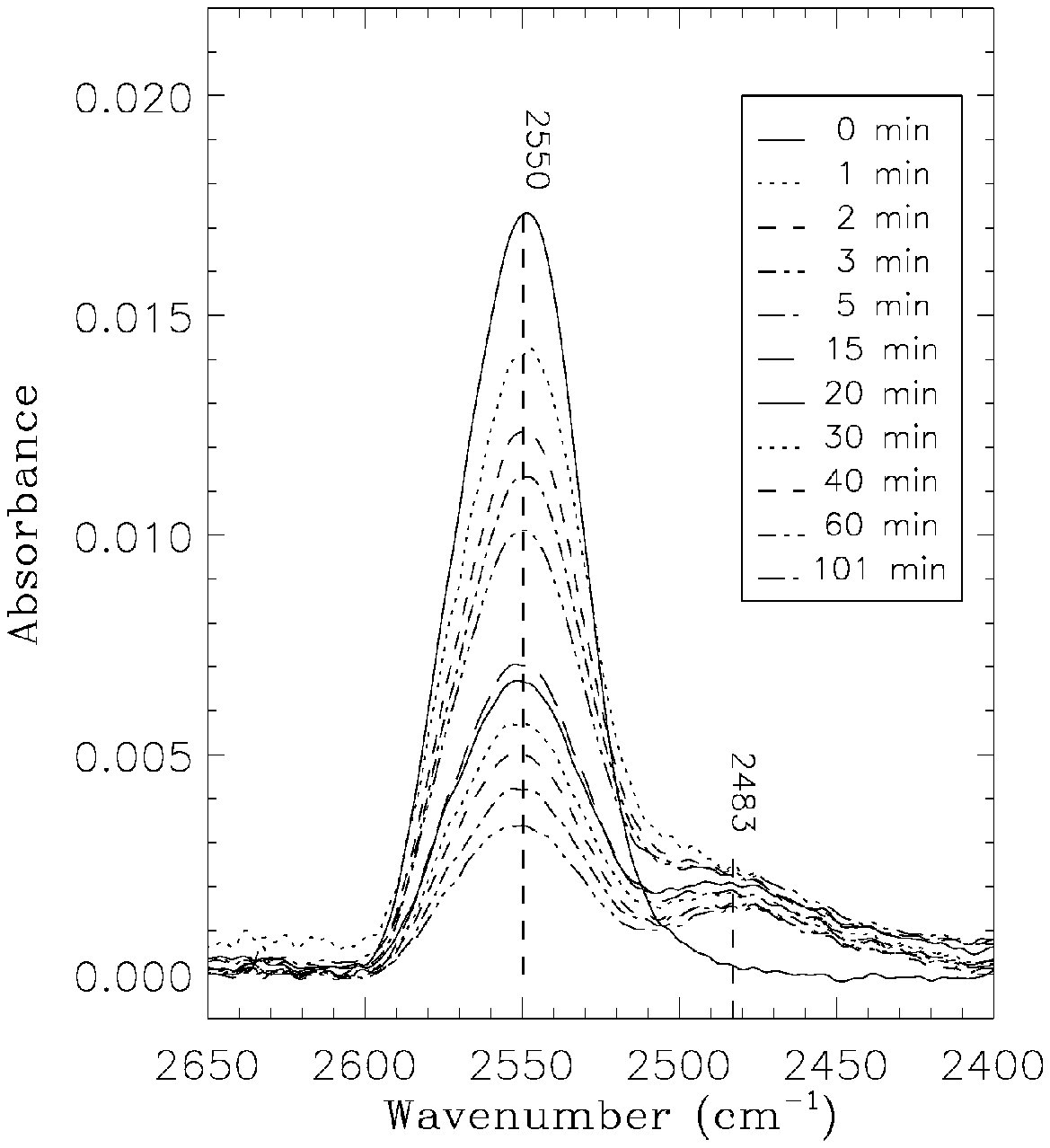} \\ 
    \includegraphics[width=7.cm]{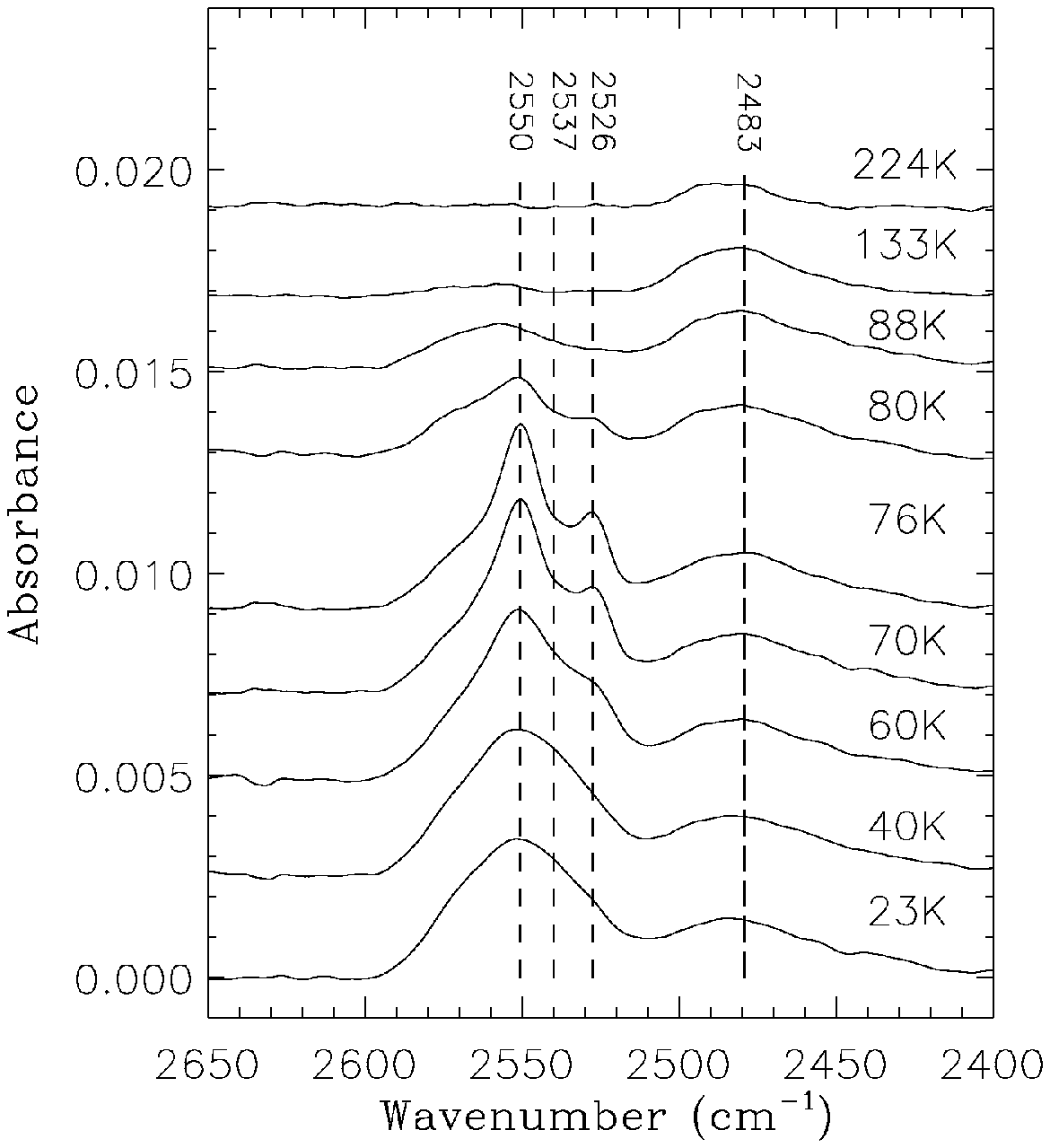} \\

   \end{tabular}
         
        \caption{Top: Infrared spectra of H$_2$S ice at 7 K for different
                 irradiation 
                 times, corresponding to experiment S8 of Table~\ref{logIR}.
                 Bottom: For the same experiment, infrared spectra of the irradiated sample collected 
                 at different temperatures during warm-up.}
        \label{ir_irrad_fig}

\end{figure}

Upon photolysis of H$_2$S, the formation of HS$^{\cdot}$ radicals occurs, 
following the reaction
\begin{equation}
{\rm H}_2{\rm S} + h \nu \stackrel{k_1}{\rightarrow} {\rm HS}^{\cdot} + {\rm H}^{\cdot}. 
\label{reaction_k1}
\end{equation}
Two radical-radical reactions are then possible:
\begin{equation}
{\rm HS}^{\cdot} + {\rm H}^{\cdot} \stackrel{k_2}{\rightarrow} {\rm H}_2{\rm S}, 
\label{reaction_k2}
\end{equation}
reforming the H$_2$S molecule or
\begin{equation}
{\rm HS}^{\cdot} + {\rm HS}^{\cdot} \stackrel{k_3}{\rightarrow} {\rm H}_2{\rm S}_2.
\label{reaction_k3}
\end{equation}
Further irradiation leads to the photolysis of the reformed H$_2$S. According to Isoniemi et al. (1999) which used a 266 nm ND:YAG laser as UV source to irradiate H$_2$S$_2$ in solid Ar, the following reactions can occur:
\begin{equation}
{\rm H}_2{\rm S}_2 + h \nu \stackrel{k_4}{\rightarrow} {\rm HS}_2^{\cdot} + {\rm H}^{\cdot}, \\
\label{reaction_k4}
\end{equation}

\begin{equation}
{\rm H}_2{\rm S}_2 + h \nu \stackrel{k_5}{\rightarrow} {\rm S}_2 + {\rm H}_2,  \\
\label{reaction_k5}
\end{equation}

\begin{equation}
{\rm HS}_2^{\cdot}  + h \nu \stackrel{k_6}{\rightarrow} {\rm S}_2 + {\rm H}^{\cdot}.
\label{reaction_k6}
\end{equation}
The top panel of Fig.~\ref{ir_irrad_fig} shows the infrared spectra of H$_2$S 
ice at 7 K for different irradiation times, corresponding to experiment 
S8 of Table~\ref{logIR}. After irradiation, the sample was warmed-up at a 
rate of 1.4 K min$^{-1}$. The infrared spectra of the irradiated sample,
collected at different temperatures during warm-up, are shown in the bottom
panel of Fig.~\ref{ir_irrad_fig}. Similar to the infrared spectra of 
the experiment with no irradiation, S2 shown in Fig.~\ref{ir_fig}, the 
spectral changes observed on the irradiated sample during warm-up, see bottom
panel of Fig.~\ref{ir_irrad_fig}, are due to crystallization of the H$_2$S ice
fraction that remains after the photolysis.
According to Isoniemi et al. (1999), the reaction rate constants corresponding 
to reactions ~\ref{reaction_k4}, \ref{reaction_k5}, and ~\ref{reaction_k6} are, 
respectively, k$_4$ = 0.041 min$^{-1}$ k$_5$ = 0.035 min$^{-1}$, and 
k$_6$ = 0.006 min$^{-1}$ for irradiation with a 266 nm laser. For short irradiation times reaction ~\ref{reaction_k5} dominates the production of S$_2$, but after a certain time reaction \ref{reaction_k6} becomes more important.
 \begin{figure}[ht!]
    \centering
      \begin{tabular}{r}

     \includegraphics[width=8.5cm]{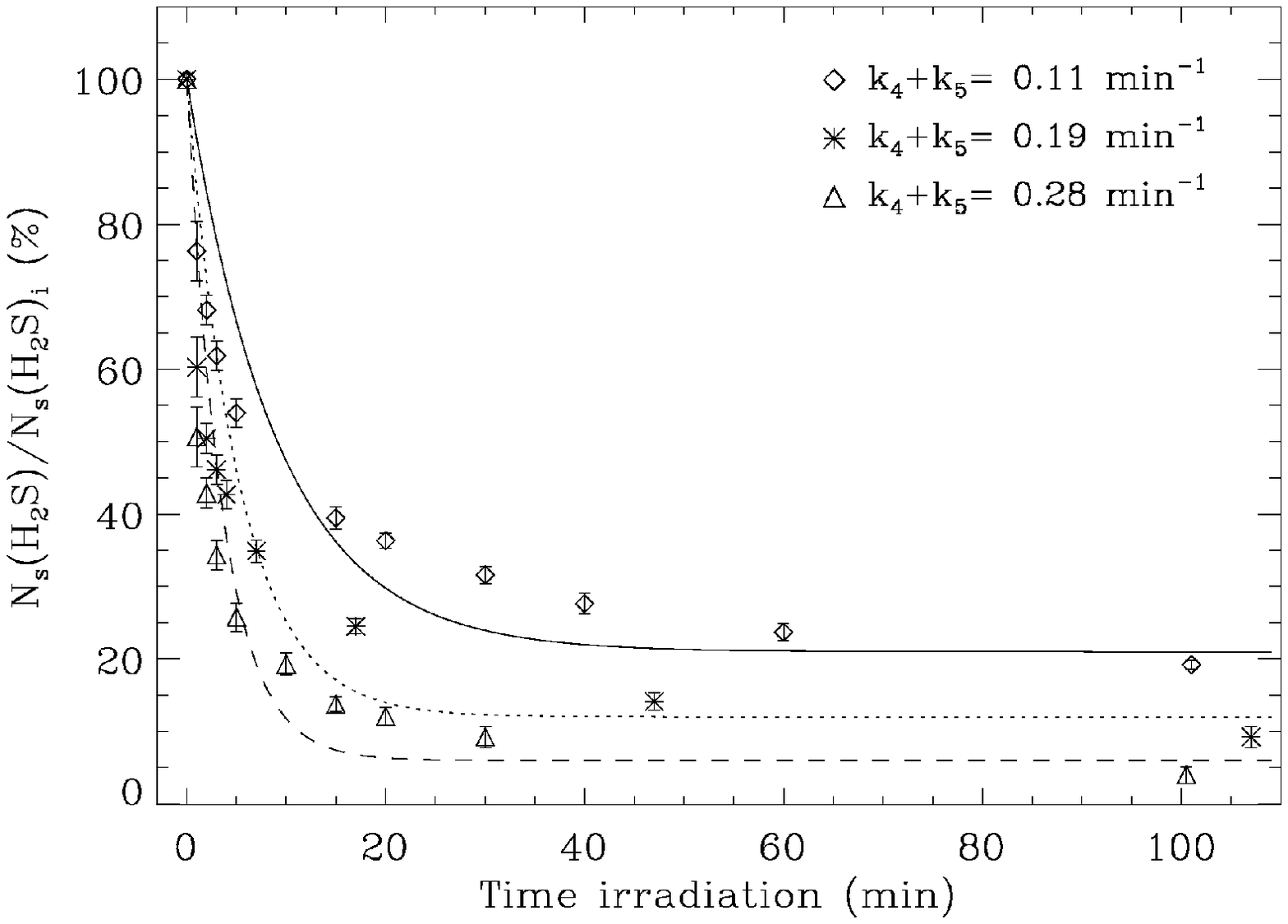}\\
    \includegraphics[width=8.5cm]{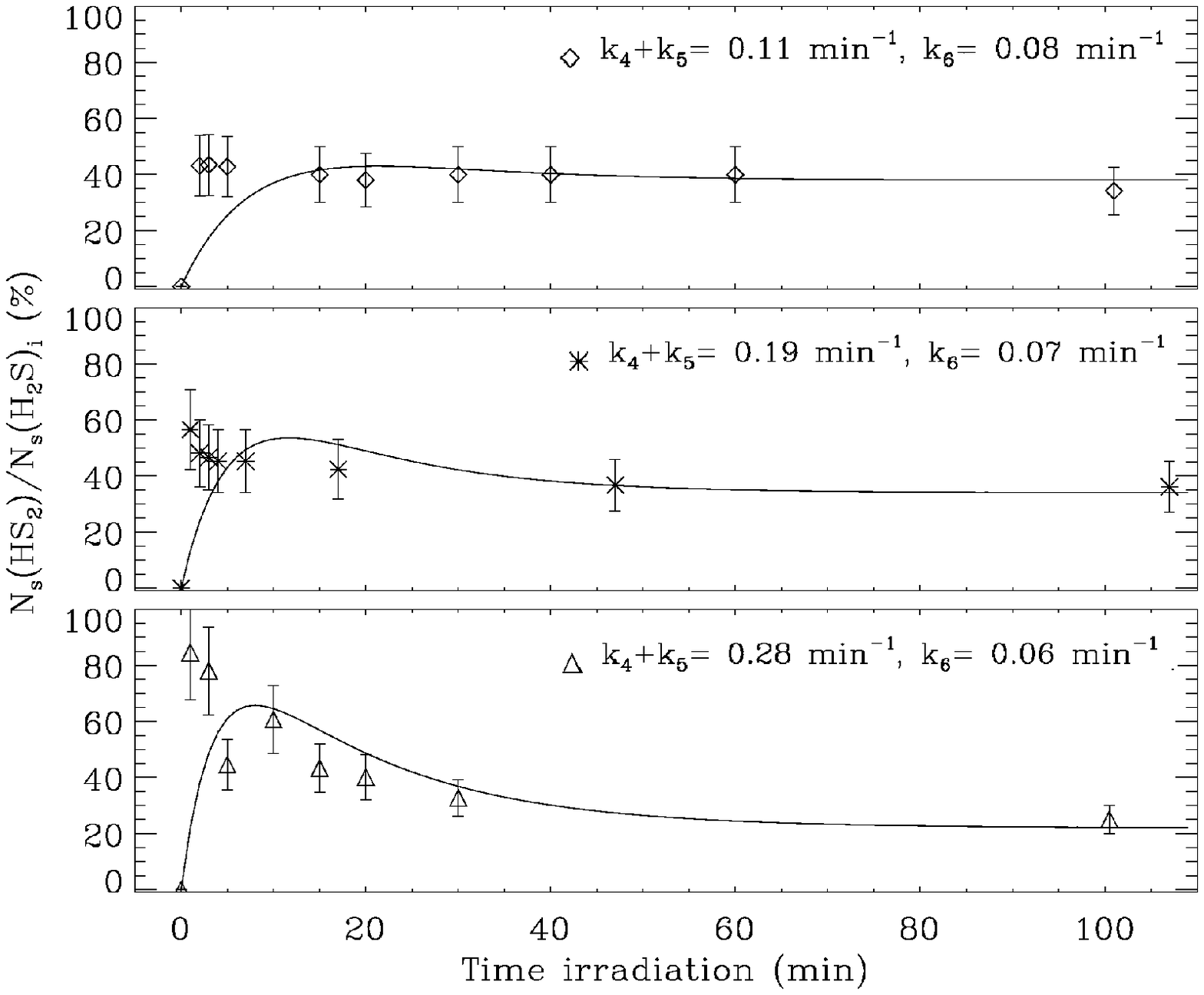} \\

   \end{tabular}
         
        \caption{Destruction of H$_2$S and formation of HS$_2^\cdot$ during UV irradiation. Top: Data points corresponding to S7 (asterisks), S8 (diamonds), and S9 (triangles) experiments, see Table~\ref{logIR}. For each experiment, the integrated absorbance values of the 2550 cm$^{-1}$ band of H$_2$S ice as a function of UV irradiation time, N$_s$(H$_2$S), were normalized with respect to their initial values before irradiation, N$_s$(H$_2$S)$_i$. The fits were made using Eq.~\ref{conc_h2s2} and rate constant values displayed in the figure. Error bars represent the standard deviation in each experiment.
Bottom: Normalized integrated absorbance values of the 2483 cm$^{-1}$ band attributed to the HS$_2^{\cdot}$ product as a function of UV irradiation time, corresponding to  S7 (asterisks), S8 (diamonds), and S9 (triangles), see Table~\ref{logIR}. Normalization was done with respect to the integrated absorbance values of the 2483 cm$^{-1}$ band after the same irradiation time. The solid line is a fit using Eq.~\ref{conc_hs2} and rate constants displayed in the figure.}
        \label{integ_abs_h2s_fig}
\end{figure}

The intricate network of reactions described above yields several 
products, with overlapping 
infrared absorption bands.
It was 
therefore not possible to distinguish between the different irradiation 
products. The absorption band centered near 2550 cm$^-$$^1$, see 
Fig.~\ref{ir_irrad_fig}, is attributed to the 
S-H stretching modes of H$_2$S, H$_2$S$_2$, and HS$^{\cdot}$, as well as their 
dimers. 
We attribute the main contribution of the neighboring band at
  2483 cm$^-$$^1$ to the S-H stretching modes of the HS$_2^{\cdot}$
  radical. Based on Isoniemi et al. (1999), other authors assign the same 2483 cm$^-$$^1$ band to
  the S-H stretching modes of H$_2$S$_2$ (Moore et al. 2007). Isoniemi
  et al. (1999) give the values of 2463 and 2460 cm$^{-1}$ for the
  HS$_2^{\cdot}$ radical, while the values assigned to H$_2$S$_2$
  are 2554 and 2557 cm$^{-1}$ in Ar matrix. The band position of
  different complexes of H$_2$S$_2$ are above 2499 cm$^{-1}$ and
  therefore the 2483 cm$^{-1}$ band is not due to
  H$_2$S$_2$. Comparing our data with that of Isoniemi et al. (1999), the S-H stretching modes of H$_2$S$_2$ are contained in the 2550 cm$^{-1}$ band. Therefore, we consider that the most plausible carrier of the 2483 cm$^{-1}$ band is the HS$_2^{\cdot}$ radical, rather than H$_2$S$_2$. These overlaps mean we could not determine the formation and destruction rate constants of all the processes involved in the irradiation of H$_2$S. We will see that our experimental data only allowed the calculation of reaction rate values k$_4$ + k$_5$, and k$_6$.
 In particular, the reformation of H$_2$S according to reaction~\ref{reaction_k2}  could not be monitored by FTIR spectroscopy.
We therefore discuss a simplified kinetics scheme that assumes that photolysis of
H$_2$S, 
yielding HS$^{\cdot}$, and subsequent formation of H$_2$S$_2$ as a first product, are 
fast processes that cannot be monitored using our FTIR spectrometer. 
 For that reason, instead of [H$_2$S]$_0$ = 1 we consider that the H$_2$S has already been converted to H$_2$S$_2$ after a short irradiation time and use
the initial conditions 
[H$_2$S$_2$]$_0$ = 1, [HS$_2^{\cdot}$]$_0$ = [S$_2$]$_0$ = 0. We then check that such approximation provides a fair fit of the experimental data. Following Isoniemi et al. (1999) we have the 
concentrations\\

\begin{equation}
[H_2S_2] = \exp{[- (k_4 + k_5) \cdot t]},
\label{conc_h2s2}
\end{equation}

\begin{equation}
[HS_2] = C \cdot \{ \exp{[- (k_4 + k_5) \cdot t]} - \exp[{- k_6 \cdot t}] \},
\label{conc_hs2}
\end{equation}

\begin{equation}
[S_2] = 1 - [H_2S_2] - [HS_2].
\label{conc_s2}
\end{equation}
   
The condition [S$_2$]$_0$ = 0 implies that 
$\frac{d [{\rm S}_2]}{dt} {\arrowvert}_0$ = 0.
Substitution of Eqs.~\ref{conc_h2s2} and \ref{conc_hs2} in 
Eq. \ref{conc_s2}, taking the first time derivative on both sides leads to 
$ C = \frac{k_4 + k_5}{k_6 - k_4 - k_5}$ in Eq.~\ref{conc_hs2}. The numerator 
of $C$, $k_4 + k_5$, differs from that given by Isoniemi et al. (1999), $k_4$.  
But our expression of $C$ is obtained analytically as explained above and should be 
correct. Eq.~\ref{conc_h2s2} was used to fit the data presented in the top panel of Fig.~\ref{integ_abs_h2s_fig}, and Eq.~\ref{conc_hs2} to fit the bottom panel of  Fig.~\ref{integ_abs_h2s_fig}. These fits lead to $k_4 + k_5$ = 0.19, 0.11, and 0.28 min$^{-1}$ and $k_6$ = 0.07, 0.08, and 0.06 min$^{-1}$ for experiments S7, S8, and S9, respectively.  
The data points in the top panel of Fig.~\ref{integ_abs_h2s_fig} are reasonably well fitted if we consider that the initial destruction of H$_2$S leading to H$_2$S$_2$ formation is not taken into account in our simplified kinetics scheme. In addition, refractory residue formation was also not considered. 

The top panel of Fig.~\ref{integ_abs_h2s_fig} also displays how the rate constants are linked to the ice thickness, showing that the largest destruction (of 96\%) corresponds to the thinnest ice, S9, while the lowest destruction (of 82\%) corresponds to the thickest ice, S8, see Table~\ref{logIR}.
The same process (k$_4$ + k$_5$) can also be monitored by following the formation of 
HS$_2^{\cdot}$ during UV irradiation of H$_2$S. This is shown in the bottom panel of Fig.~\ref{integ_abs_h2s_fig} where the abundances of HS$_2^{\cdot}$ with respect to H$_2$S, associated to the 2483 cm$^{-1}$ band are displayed.
 
However, additional elongation reactions of HS$_2^\cdot$ are probably present and compete with S$_2$ formation, elongating the HS$_2^\cdot$ molecule up to H$_2$S$_8$ and finally forming S$_8$. The polymerization process is favored by the high N$_s$(HS$_2^\cdot$)/N$_s$(H$_2$S) ratio in the first minutes of irradiation encouraging the reaction between two HS$_2^\cdot$ molecules. This could explain the higher abundance of HS$_2^\cdot$ in the first stages of irradiation with respect to the final irradiation time for the thinnest ice, which has the lowest rate (k$_6$ = 0.06 min$^{-1}$). 

\begin{figure}[ht!]
    \centering
      \begin{tabular}{r}

    \includegraphics[width=7.cm]{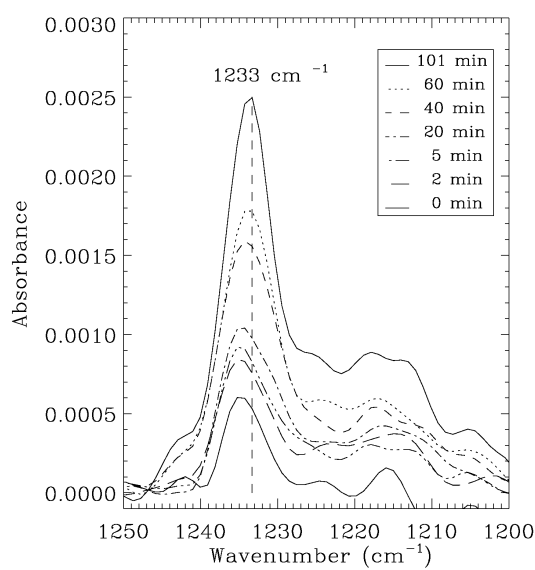} \\ 

   \end{tabular}
         
        \caption{Infrared band around 1233 cm$^{-1}$ formed by UV irradiation of H$_2$S 
        ice at 7 K, corresponding to experiment S8 of Table~\ref{logIR}. Spectra were collected for 
        different irradiation times, see inlet. This band is attributed to H$_2$SS based on calculations by Isoniemi et al. (1999).}
       \label{H2SS_1_fig}
\end{figure}

\begin{figure}[ht!]
    \centering
      \begin{tabular}{r}

    \includegraphics[width=8.cm]{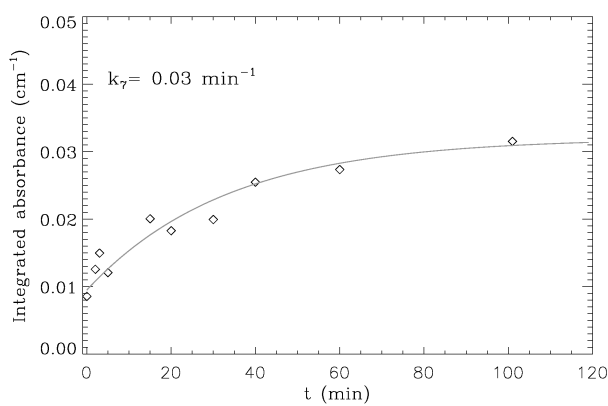} \\ 
   
   \end{tabular}
         
        \caption{Integrated absorbance values of the 1233 cm$^{-1}$ band 
        attributed to H$_2$SS ice as a function of UV irradiation time, 
        corresponding to 
        experiment S8, see Table~\ref{logIR}. The solid line is a first-order fit using rate constant k = 0.03 min$^{-1}$, which is probably the value of k$_7$ from Eq.~\ref{reaction_k7}.}
       \label{H2SS_2_fig}
\end{figure}

H$_2$SS might also be a product of H$_2$S ice 
photoprocessing in our experiments. It has a greater energetic level than its H$_2$S$_2$ isomer, roughly 160 
kJ mol$^{-1}$ larger (Gerbaux et al. 2000). Nevertheless, H$_2$SS has a lower 
stability, due to its energetic sizable barrier that prevents  
rearrangement back to the more stable H$_2$S$_2$ form at cryogenic 
temperatures. Thermal isomerization of H$_2$S$_2$ leading to H$_2$SS can therefore be excluded, and H$_2$SS should be kinetically stable toward 
unimolecular isomerization at low temperatures (Steudel et al. 1997). For this 
reason, only a photochemical process can explain the formation of H$_2$SS. 
Two different reactions can generate H$_2$SS:

\begin{equation}
{\rm H}_2{\rm S} + {\rm S}^{\cdot} \stackrel{k_7}{\rightarrow} {\rm H}_2{\rm SS} , \\
\label{reaction_k7}
\end{equation}
and
\begin{equation}
{\rm H}{\rm S}_2^{\cdot} + {\rm H}^{\cdot} \stackrel{k_8}{\rightarrow} {\rm H}_2{\rm SS}. \\
\label{reaction_k8}
\end{equation}

The infrared frequencies of H$_2$SS have
been calculated by different methods, where the 
$\nu$$_2$(HSH-bending) of H$_2$SS is expected to fall at 1236 cm$^{-1}$ 
(Isoniemi et al. 1999). During UV irradiation of H$_2$S ice, a new band grows well above the noise level at 1233 cm$^-$$^1$ with a signal-to-noise ratio of 3.7. This band is shown for different irradiation 
times in Fig.~\ref{H2SS_1_fig}, and the corresponding 
integrated absorbance values are shown in Fig.~\ref{H2SS_2_fig}.
The solid line is the fit, giving k$_7$ = 0.03 min$^{-1}$. The position and the growth of this band as a function of irradiation time suggests that its molecular carrier is H$_2$SS.

\begin{figure}[ht!]
    \centering
      \begin{tabular}{r}

    \includegraphics[width=9.cm]{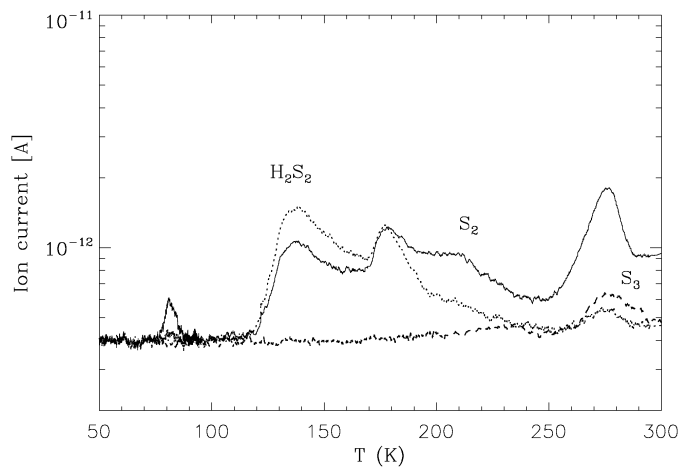} \\ 
    
   \end{tabular}
         
        \caption{Thermal desorption of UV-irradiated H$_2$S ice, corresponding 
        to experiment S8 of Table~\ref{logIR}. Spectra correspond to m/z = 66 (dotted trace), m/z = 64 (solid trace), m/z = 96 (dashed trace), the molecular 
        ion masses of the H$_2$S$_2$, S$_2$, and S$_3$ products. The ion current in Ampere, represented on the y-scale, corresponds roughly to the partial pressure in mbar.}
       \label{polimeros_S_fig}
\end{figure}
UV irradiation of H$_2$S ice leads to the formation of S-polymers, which were 
detected at room temperature by chromatographic techniques (Mu\~noz Caro 2002), see 
Sect.~\ref{intro}. Figure~\ref{polimeros_S_fig} shows the TPD spectra 
corresponding to the H$_2$S$_2$, S$_2$, and S$_3$ products, measured during the warm-up of experiment S8, see Table~\ref{logIR}. 
The peaks at 178 and 138 K in the m/z = 66 spectrum may be due to the cis- and trans- isomers of H$_2$S$_2$, which could be formed by reclustering of HS$_{2}^{\cdot}$ during warm-up.
The desorption temperature of the cis-isomer is higher because its packing energy is greater than that of the trans-isomer, although the former has less stability than the latter (C\'ardenas-Jir\'on et al. 1990).
The lower stability of the cis-isomer is a more prominent steric effect, because their hydrogens are partially eclipsed. Alternatively, the same peaks at 178 and 138 K could be attributed to a mixture of H$_2$S$_2$ and H$_2$SS.

The peaks at 212 and 275 K most likely correspond to S$_2$, and S$_3$, 
respectively.
The formation of S$_2$ and S$_3$ in the ice cannot be observed by mid-infrared 
spectroscopy. Addition of S$^{\cdot}$ or SH$^{\cdot}$ to other species such as 
H$_2$S$_2$, HS$_{2}^{\cdot}$, and S$_{2}^{\cdot}$ lead to formation of S$_3$ and 
other S-polymers up to S$_8$ (Barnes et al. 1974). 
\subsection{Irradiation of H$_2$O:H$_2$S ice experiments}
\label{irrad_2}

\begin{figure}[ht!]
    \centering
      \begin{tabular}{r}

      \includegraphics[width=7cm]{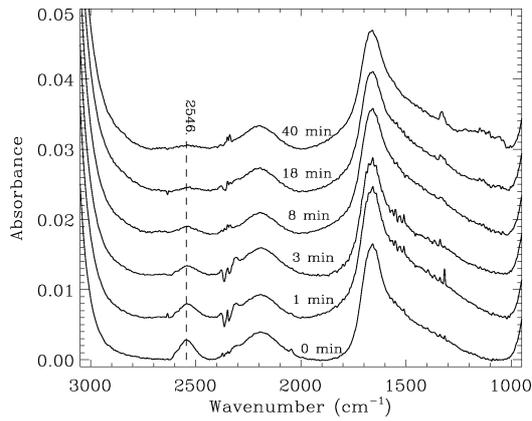} \\

   \end{tabular}
         
        \caption{Infrared spectra of the H$_2$S:H$_2$O = 4.0:100 ice mixture at 8 K for
        different irradiation times, corresponding to experiment S10 of Table 1.}
       \label{irrad_mixture}
\end{figure}

The H$_2$S:H$_2$O = 4.0:100 ice mixture was irradiated at 8 K, see experiment
S10 of Table 1 for experimental parameters. The corresponding infrared spectra 
for different irradiation times are displayed in Fig.~\ref{irrad_mixture}. 
While the H$_2$S band at 2546 cm$^{-1}$ decreases as a function of 
irradiation time, two new bands appear at 1328 and 1151 cm$^{-1}$, and the latter 
is very weak, which correspond to the asym. and sym. str. modes of SO$_2$.

\begin{figure}[h!]
    \centering
      \begin{tabular}{r}

   \includegraphics[width=8.cm]{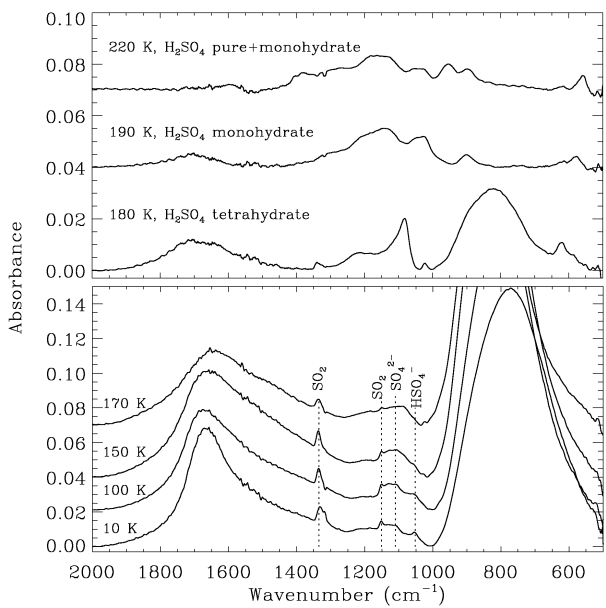} \\ 
   
   \end{tabular}

        \caption{Infrared spectra of the H$_2$S:H$_2$O = 13:100 ice mixture  for different temperatures deposited at 8 K, corresponding to experiment S12 of Table 1. The 
         deposition and irradiation was performed simultaneously in this 
         experiment.}
       \label{irrad_mixture_2}
\end{figure}

Other weak bands appeared in the 1300--1000 cm$^{-1}$ region. To
characterize these bands better we performed a simultaneous deposition and irradiation experiment with the less 
diluted H$_2$S:H$_2$O = 13:100 ice mixture, experiment S12 of Table~\ref{logIR}.
The infrared spectra of the irradiated sample, collected at different 
temperatures during warm-up, are shown in Fig.~\ref{irrad_mixture_2}. 
The SO$_2$ bands are also observed in this experiment. Two new bands near 
1110 and 1052 cm$^{-1}$ are associated to SO$_4^=$ and HSO$_4^{-}$, based on
previous assignments by Kunimatsu et al. (1988) and Moore et al. (2007).
These species correspond to higher oxidation states of photoprocessed SO$_2$. 

Upon warm-up the band of HSO$_4^{-}$ at 1052 cm$^{-1}$ decreases and the 
SO$_4^=$ band at 1110 cm$^{-1}$ shifts to 1083 cm$^{-1}$ at 180 K, when 
its intensity is maximum. After water desorption, H$_2$SO$_4$ tetrahydrate 
formed at 180 K. At 190 K new absorption bands appear at 1142, 1024, and 
901 cm$^{-1}$. These absorption bands indicate the presence of 
H$_2$SO$_4$ monohydrate. At 220 K the bands observed at 1374 and 954 cm$^{-1}$ 
can be interpreted as a mixture of H$_2$SO$_4$ in its pure state and 
H$_2$SO$_4$ monohydrate, following the Moore et al. (2007) assignment. 
This ejection of water continues up to room temperature (Couling et al. 2002).

\begin{figure}[ht!]
    \centering
      \begin{tabular}{r}

    \includegraphics[width=8.cm]{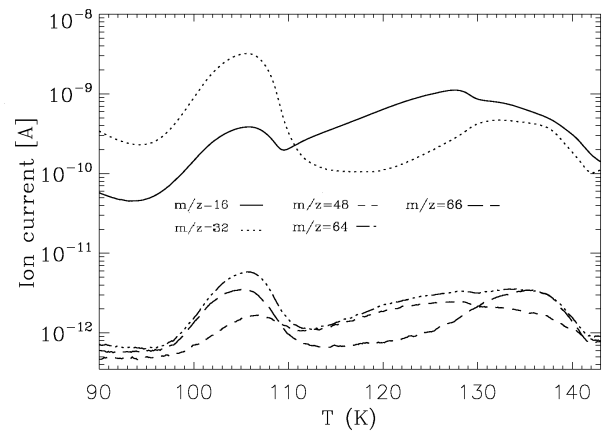} \\ 
   
   \end{tabular}
         
        \caption{Thermal desorption of UV-irradiated H$_2$S:H$_2$O = 13:100
        in the range of 90--140 K, corresponding to experiment
        S12 of Table 1. The ion current in Ampere, represented on the y-scale, 
        corresponds roughly to the partial pressure in mbar.}
       \label{irrad_mixture_3}
\end{figure}

TPD curves of H$_2$S:H$_2$O = 13:100 ice in the 90--143 K range are shown in 
Fig.~\ref{irrad_mixture_3}. A desorption peak present at 106 K is common to the 
m/z = 16, 32, 48, 64, and 66 spectra. These mass fragments correspond to 
H$_2$SO$_2$ or a mixture of H$_2$SO$_2$ and SO$_2$. The H$_2$SO$_2$ molecule 
has not been previously detected, but it is thought to be necessary as an 
oxidation intermediate to yield oxides with elevated oxidation states 
(Steiger \& Steudel 1992).  
Desorption of SO$_2$ at 128 K is observed as an increase in the
m/z = 16, 48, 64 spectra. Overlapping with the desorption of SO$_2$ is 
another peak around 137 K common to m/z = 32, 64, 66. This desorption is 
assigned to H$_2$S$_2$, which desorbed at a similar temperature, 138 K, in the
pure H$_2$S irradiation experiment, see Fig.~\ref{polimeros_S_fig}.
Two of the above products, H$_2$S$_2$ and SO$_2$, were detected by infrared 
spectroscopy after proton bombardment of the H$_2$O:H$_2$S = 8:1 ice mixture; 
the other products detected by us in the H$_2$S:H$_2$O UV irradiation experiments, were formed by proton bombardment of
H$_2$O:SO$_2$ ice mixtures (Moore et al. 2007). 

\begin{figure}[h!]
    \centering
      \begin{tabular}{r}
    \includegraphics[width=8.cm]{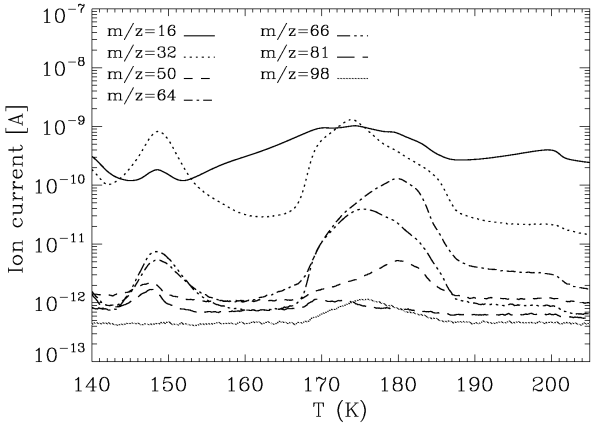} \\ 
   
   \end{tabular}
         
        \caption{Thermal desorption of UV-irradiated H$_2$S:H$_2$O = 13:100 
        in the 140--205 K range, corresponding to experiment
        S12 of Table 1. The ion current in Ampere, represented on the y-scale, corresponds roughly to the partial pressure in mbar.}
       \label{irrad_mixture_5}
\end{figure}

The TPD curves of the same H$_2$S:H$_2$O = 13:100 ice irradiation 
experiment in the 140--205 K range are displayed in Fig.~\ref{irrad_mixture_5}.
A desorption peaking at 149 K is common to m/z= 16, 32, 64, 66, and 81. 
This combination of mass fragments corresponds to a possible desorption of 
the HSO$_3^-$ anion. No desorption of mass fragment m/z= 82, corresponding to
H$_2$SO$_3$, was observed. 

Both infrared and QMS data, Figs.~\ref{irrad_mixture_2} and \ref{irrad_mixture_5}, show
that a large fraction of the irradiation products mentioned above is retained in the 
H$_2$O-matrix and finally co-desorb with H$_2$O between 160--190 K. Also a fraction of the
more refractory H$_2$SO$_4$, m/z = 98, co-desorbs with H$_2$O around 175 K, see 
Fig.~\ref{irrad_mixture_5}. 

\begin{table*}[ht!]
\centering
\caption{Parameters of the experiments.}
\label{logIR}
\begin{tabular}{cccccccccc}   \hline \hline
Exp. & Comment & Ice mixture & Fluence & Heating rate & N(H$_2$S) & $\frac{{\rm N}_{\rm S}{\rm (HS_2^{\cdot})}}{{\rm N}_{\rm S}{\rm (H_2S)}}$ & 1-$\frac{{\rm{ N}}_{\rm{ S}}{\rm{ (H_2S)}}}{{\rm{ N}}_{\rm{ S}}{\rm{ (H_2S)_i}}}$ \\ 
S\#  &         & \tiny{H$_2$S:H$_2$O} & $\frac{\rm {photons}}{\rm {cm^2}}$ & $\frac{\rm {K}}{\rm {min}}$ & cm$^{-2}$ & (\tiny\%)  & (\tiny\%) \\  \hline   
1   & Deposition & 1:0     & - & 1.1 & 1.2 $\times$ 10$^{16}$ & - & -  \\ 
2   & Deposition & 1:0     & - & 0.5 & 6.8 $\times$ 10$^{16}$ & - & -  \\
3   & Deposition & 1:0     & - & 1.3 & 8.4 $\times$ 10$^{16}$&- & - \\ 
4  & Deposition & 1:0   & - & 1.0 & 6.8 $\times$ 10$^{16}$& - & -  \\ 
5   & Deposition & 13.7:100   & - & 2.0 & 1.5 $\times$ 10$^{16}$ $^a$& - & - \\ 
6  & Deposition & 7.5:100   & - & 1.0 & 1.3 $\times$ 10$^{16}$ $^a$& - & -  \\ 
7   & Dep., after irrad. & 1:0     & 1.6 $\times$ 10$^{18}$ & 3.3 & 7.0 $\times$ 10$^{16}$& 36.($\pm$9) & 90($\pm$1) \\ 
8   & Dep., after irrad. & 1:0     & 1.5 $\times$ 10$^{18}$ & 1.4 & 9.9 $\times$ 10$^{16}$ & 34.($\pm$8) & 82($\pm$1) \\ 
9   & Dep., after irrad. & 1:0     & 1.5 $\times$ 10$^{18}$ & 2.1 & 3.7 $\times$ 10$^{16}$ & 25.($\pm$6) & 96($\pm$1) \\ 
10  & Dep., after irrad. & 4.0:100  & 6.0 $\times$ 10$^{17}$  & 1.0 & 3.0 $\times$ 10$^{16}$ $^a$& - & 75($\pm$1) \\ 
12   & Simultaneus dep. and irrad. & 13:100 & 3.8 $\times$ 10$^{18}$ & 1.0  & 3.8 $\times$ 10$^{17}$ & - & -  \\
\hline
\end{tabular}
\begin{list}{}
\item  $^a$ Column density of H$_2$S obtained from integration of infrared absorption assuming $A$(H$_2$S)$_{pure}$ = $A$(H$_2$S)$_{mixed}$ = 2.0 $\times$ 10$^{-17}$ cm molec$^{-1}$.  
\end{list}
\end{table*}

\section{Astrophysical implications}
\label{astro}

Comparison of our laboratory spectra with observations of young stellar 
objects (YSOs), performed with the ISO satellite, show an absorption band 
compatible with the 2548 cm$^{-1}$ (3.925 $\mu$m) feature of solid H$_2$S 
in the ice mantles. The absorbances toward different YSOs were calculated as follows
\begin{equation}
 Absorbance =- log_{10}(\frac{I}{I_0})
\label{abs}
\end{equation}
where $I$ is the flux, in Jy, measured toward the YSO line of sight, and $I_0$ is the continuum flux. 
\begin{figure}[ht!]
    \centering
      \begin{tabular}{r}

       \includegraphics[width=8.5cm]{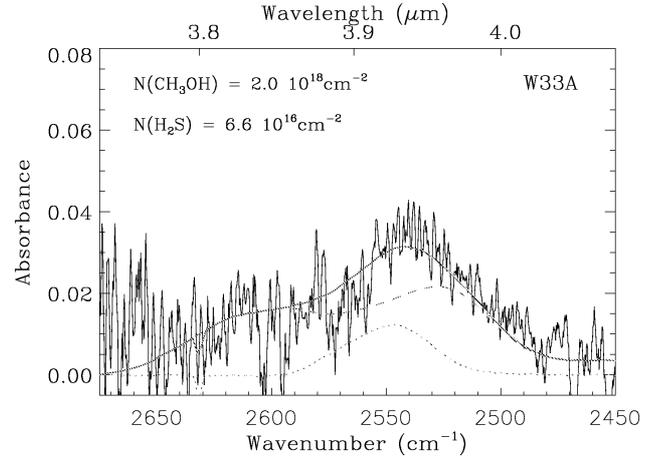} \\
       \end{tabular}
         
        \caption{Comparison between our experimental data and the ISO 
        observation of W33A. Dashed line corresponds to the pure CH$_3$OH 
        ice spectrum at 20 K, dotted line corresponds to pure H$_2$S ice 
        at 20 K, and the continuous line is the addition of both spectra.}
       \label{w33a}
\end{figure}

Methanol ice displays two weak bands at 2530 and 2610 cm$^{-1}$ 
(3.95 and 3.83 $\mu$m), which are assigned to the combination modes (Dartois et al. 1999), and can overlap with the H$_2$S 
absorption at 2548 cm$^{-1}$ (3.925 $\mu$m). The fit of the 3.95 $\mu$m feature 
with the laboratory spectrum of pure CH$_3$OH allows setting an upper limit on 
the 3.83 $\mu$m CH$_3$OH absorption. In addition, the $\sim$ 18 cm$^{-1}$ 
(0.025 $\mu$m) difference in the position of the H$_2$S feature at 
3.925 $\mu$m with respect to the CH$_3$OH feature at 3.95 $\mu$m also allows 
 differentiation of both absorptions. It is therefore possible to 
provide an upper limit on the value of N(H$_2$S).   

The column density of CH$_3$OH toward W33A, N(CH$_3$OH), which is obtained from integration of the 3.95 and 3.83 $\mu$m bands, is about 2 $\times$ 10$^{18}$ cm$^{-2}$ (Dartois et al. 1999) from UKIRT observations.
 The different values of N(CH$_3$OH) depend on the infrared band used for integrating the absorbance and the telescope used. For instance, Allamandola et al. (1992) obtain N(CH$_3$OH) = 4.0 $\times$ 10$^{18}$ cm$^{-2}$ from integration of the 2825 cm$^{-1}$ (3.53 $\mu$m) band toward W33A using the 3 m telescope at IRTF.  
Figure~\ref{w33a} shows the spectrum of protostar W33A, which is similar to what is reported by Teixeira et al. (1999) from ISO observations. The laboratory fit consists of adding pure CH$_3$OH and pure H$_2$S ice spectra, both at 20 K. We obtained N(CH$_3$OH) = 2.0 $\times$ 10$^{18}$ cm$^{-2}$ and 
N(H$_2$S) $\le$ 6.6 $\times$ 10$^{16}$ cm$^{-2}$, as shown in Table~\ref{ratios}.

\begin{figure}[ht!]
    \centering
      \begin{tabular}{r}

       \includegraphics[width=8.cm]{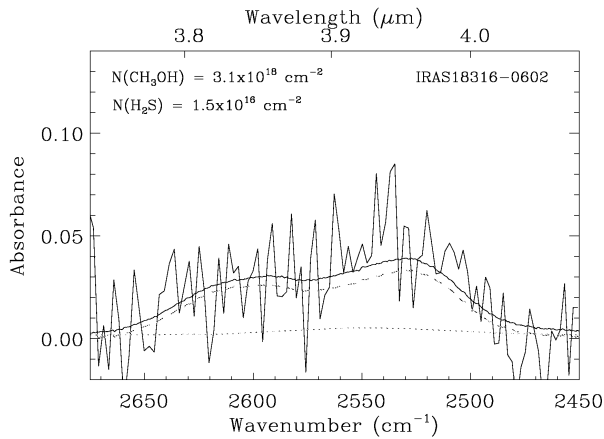} \\ 
   
   \end{tabular}
         
        \caption{Comparison between our experimental data and the ISO 
        observation 
        of IRAS18316-0602. Dashed line corresponds to the pure CH$_3$OH 
        ice spectrum at 20 K, dotted line corresponds to the 
        H$_2$S:H$_2$O= 13.7:100 ice mixture at 20 K, and continuous line 
        is the addition of both spectra.}
       \label{iras}
\end{figure}

The best fit of protostar IRAS18316-0602 spectrum, shown in 
Fig.~\ref{iras}, is obtained by adding the spectrum of pure CH$_3$OH at 20 K
with that of the H$_2$S:H$_2$O = 13.7:100 ice mixture at 20 K. The column density
of pure CH$_3$OH toward this source is 
N(CH$_3$OH) = 3.8 $\times$ 10$^{18}$ cm$^{-2}$ (Dartois et al. 1999) using the UKIRT telescope.
\begin{table*}
\centering
\caption{Observed ratios toward protostars.}
\label{ratios}
\begin{tabular}{cccccccccc}   \hline \hline
Source & N(CH$_3$OH) & N(H$_2$S) & N(H$_2$O) &  $\frac{\rm N(H_2S)}{\rm N(H_2O)}$\\
 & (cm$^{-2}$)  & (cm$^{-2}$)& (cm$^{-2}$)& (\%) \\ \hline
W33A & 2.0 $\times$ 10$^{18a}$ & $\le$ 6.6 $\times$ 10$^{16a}$ & 0.9-4.2 $\times$ 10$^{19b}$ & $\le$ 0.7-0.2$^a$\\
     &  2 $\times$ 10$^{18b}$   &      &     & \\ \\
IRAS18316-0602 & 3.1 $\times$ 10$^{18a}$ & $\le$ 1.5 $\times$ 10$^{16a}$ & 1.2 $\times$ 10$^{19b}$ & $\le$ 0.13$^a$ \\
               &   3.8 $\times$ 10$^{18b}$    & & &\\
              
\hline

\end{tabular}
\begin{list}{}
\item \ \ \ \ \ \ \ \ \ \ \ \ \ \ \ \  \ \ \ \ \ \ \ \ \ \ \ \ \ \ \ \ \ \ \ \ \ \ \ \ $^a$ This work. $^b$ Dartois et al. 1999. 
\end{list}
\end{table*}

The upper limits of 0.7\% and 0.13\% for the H$_2$S abundances given in 
Table~\ref{ratios}, corresponding to W33A and IRAS18316-0602, are well below 
the H$_2$S/H$_2$O = 1/37
abundance ratio (corresponding to 2.7\% H$_2$S relative to H$_2$O) based on the 
S/O = 1/37 cosmic abundance ratio, see Sect.~\ref{intro}. 
The products formed by UV irradiation 
of H$_2$S in ice mantles might also account for the missing sulfur in dense 
clouds and circumstellar regions. The more refractory products could stick to 
the grain even after sublimation of the ice mantle. If we assume a flux 
value of 10$^4$ photons cm$^{-2}$ s$^{-1}$ in a dense cloud interior 
(Shen et al. 2004 and ref. therein), ice mantles experience a fluence of 
3.2 $\times$ 10$^{17}$ photons cm$^{-2}$
after 10$^6$ yr, which using Eq. ~\ref{Abs_cross} with $\sigma$ $\approx$ $\sigma$(H$_2$O) = 4.0 $\times$ 10$^{-18}$ cm$^2$ corresponds to 12\% absorption or 3.8 $\times$ 10$^{16}$ photons absorbed cm$^{-2}$. If we divide the number of absorbed photons cm$^{-2}$ by the ice column density of 3.12 $\times$ 10$^{16}$  molecules cm$^{-2}$, i.e a 0.01 $\mu$m thick ice mantle we find that roughly, each molecule absorbed 1.2 photons on average.
Ideally, a relevant laboratory simulation of such a process should use the same ice thickness of 0.01 $\mu$m, but a quantitative estimation of the irradiation products in this ice is not possible in practice. As a rough approximation, we made an experiment where the average number of photons absorbed per molecule is similar.
Experiment S10 is the closest to those conditions. It involved the irradiation of H$_2$S:H$_2$O = 4.0:100 ice, so it was exposed to $\approx$ 0.8 photon molecule$^{-1}$. 
About 75 \% of the H$_2$S molecules in 
the ice were photoprocessed, leading to the formation of products in the experiment. Among the H$_2$S-photoprocessing products that 
could be present in ice mantles, H$_2$S$_2$ and HS$^{\cdot}$ are difficult to
detect because their absorptions overlap with the H$_2$S band near 
2548 cm$^{-1}$ (3.925 $\mu$m). The neighboring band at 2483 cm$^{-1}$ (4.027 $\mu$m),
attributed to the HS$_2^{\cdot}$ and H$_2$SS species, see Sect.~\ref{irrad} and 
top panel of Fig.~\ref{ir_irrad_fig}, is expected to be below the detection 
limit of current infrared observations. 
With the exception of the SO$_2$ detection (Boogert et al. 1997) in circumstellar ice mantles, the oxidized S species, formed by irradiation of H$_2$S:H$_2$O ice mixtures, Sect.~\ref{irrad_2}, have so far not been detected by infrared spectroscopy, although SO$_2$ species has been detected in the solid phase.
Other products of 
H$_2$S-photoprocessing are the S-polymers, from S$_2$ to S$_8$, and organic 
species containing S (Mu\~noz Caro 2002). Atomic S was detected by ISO at 
25.2 $\mu$m in the Orion H$_2$ emission peak
(Rosenthal et al. 2000). Rotational transitions of sulfur chains will be 
observable with ALMA (Wakelam et al. 2005). Cyclic S$_8$ is active in the 
infrared region, but has 
not been observed in the emission spectra of comets. It could be detected by 
simple {\it in situ} GC-MS analysis, i.e. without derivatization, of a comet 
nucleus planned by the ongoing Rosetta mission (Mu\~noz Caro 2002; Goesmann et al. 2007).
%
\section{Conclusions}
\label{conc}
We determined the desorption temperature of H$_2$S ice, which depends on the 
initial H$_2$S/H$_2$O ratio. Pure H$_2$S ice desorbs around 82 K. When H$_2$S is present in an H$_2$O ice matrix, a fraction
of the H$_2$S co-desorbs with H$_2$O in the 130--170 K temperature range, showing two maxima around 145 and 163 K. These results agree with Collings et al. (2004).
Comparison of the laboratory infrared spectra of H$_2$S, pure or in an 
H$_2$O matrix, with ISO observations of protostars W33A and IRAS18316-0602
provided upper limits of 0.7 and 0.13\% on the solid H$_2$S abundance relative
to H$_2$O. These values are too low to explain the S depletion observed toward 
dense clouds and circumstellar regions. Another reservoir of S in these regions
could be the products of H$_2$S ice photoprocessing.   
It was found that solid H$_2$S, which expected in icy grain mantles, 
photolyzes very readily, leading to the formation of several species, including
H$_2$S$_2$, HS$^{\cdot}$, HS$_2^{\cdot}$, S$_2$, and H$_2$SS. The set of 
reactions leading to these species is known, and it was possible to determine 
the values of some of the rate constants involved. If H$_2$S was present in an 
H$_2$O ice matrix the SO$_2$, SO$_4^=$, HSO$_3^-$, HSO$_4^-$, H$_2$SO$_2$, 
H$_2$SO$_4$, and H$_2$S$_2$ photoprocessing products were formed in our 
experiments. 
Proton bombardment of H$_2$S:H$_2$O ice led to formation of H$_2$S$_2$ and 
SO$_2$, while a similar processing of SO$_2$:H$_2$O ice formed 
H$_2$O$_2$, H$_3$O$^+$, HSO$_3^-$, HSO$_4^-$, and SO$_4^=$ (Moore et al. 2007).
With the exception of H$_2$SO$_4$, all the above products of H$_2$S:H$_2$O ice
photoprocessing desorbed between 100-200 K. S-polymers from S$_3$ to 
S$_8$ are also products of H$_2$S photoprocessing; these species can stick 
on grains even after sublimation of the ice mantle (Mu\~noz Caro 2002, this 
work). In general, given the expected relative low abundances of S-polymers 
in ice mantles, they would be difficult to detect by infrared spectroscopy, 
but some of them will be observable with ALMA in the gas phase 
(Wakelam et al. 2005).
The UV-irradiation of H$_2$O:CO:CH$_3$OH:NH$_3$:H$_2$S 
ices is left for future work. It was found that S$_8$ was by far the most 
abundant refractory product in these experiments, even for low initial 
H$_2$S ice abundances on the order of 2\% relative to H$_2$O ice (Mu\~noz 
Caro 2002). Including C in H$_2$S-containing ices, in the form of 
abundant ice components such as CH$_3$OH or CO, may lead to species like 
OCS, which is observed in circumstellar ice mantles. 
\begin{acknowledgements}
We acknowledge U. J. Meierhenrich and W. A. Schuttte for collaboration on 
previous experiments involving chromatographic analysis of residues made
from UV-irradiation of H$_2$S-containing ice, published in the Ph.D. thesis
of G. M. M. C.
We thank J. R. Goicoechea for his support in reducing ISO data and
J. Sobrado for technical support. We also thank the ISO database and W. Frieswijk et al. for providing the data collection and previous data processing of spectra of W33A and IRAS18316-0602.
A.J. was financed by a training grant from INTA.
G.M.M.C. was supported by a Ram\'on y Cajal research contract from the MICINN in 
Spain. This research was financed by the Spanish MICINN under Project AYA2008-06374 and CONSOLIDER grant CSD2009-00038.
\end{acknowledgements}

\end{document}